\documentclass[9pt,twocolumn,twoside]{optica}
\setboolean{shortarticle}{true}
\setboolean{minireview}{false}

\title{Crystalline optical cavity at 4 K with thermal noise limited instability and ultralow drift}

\author[1, *]{John M. Robinson}
\author[1]{Eric Oelker}
\author[1]{William R. Milner}
\author[1, $\dagger$]{Wei Zhang}
\author[2]{Thomas Legero}
\author[2, $\ddagger$]{Dan G. Matei}
\author[2]{Fritz Riehle}
\author[2]{Uwe Sterr}
\author[1]{Jun Ye}

\affil[1]{JILA, NIST and University of Colorado, 440 UCB, Boulder, Colorado 80309, USA}
\affil[2]{Physikalisch-Technische Bundesanstalt, Bundesallee 100, 38116 Braunschweig, Germany}
\affil[$\dagger$]{Current address : National Institute of Standards and Technology
325 Broadway, Boulder, CO 80305, USA}
\affil[$\ddagger$]{Current address : Horia Hulubei National Institute of Physics and Nuclear Engineering, Reactorului 30, 077125 Magurele, Romania
}

\affil[*]{Corresponding author: john.robinson@colorado.edu}

\begin{document}
\pagestyle{empty}

\begin{abstract}
Crystalline optical cavities are the foundation of today's state-of-the-art ultrastable lasers. Building on our previous silicon cavity effort, we now achieve the fundamental thermal noise-limited stability for a 6 cm long silicon cavity cooled to 4 Kelvin, reaching $6.5\times10^{-17}$ from 0.8 to 80 seconds. We also report for the first time a clear linear dependence of the cavity frequency drift on the incident optical power. The lowest fractional frequency drift of $-3\times10^{-19}$/s is attained at a transmitted power of 40 nW, with an extrapolated drift approaching zero in the absence of optical power. These demonstrations provide a promising direction to reach a new performance domain for stable lasers, with stability better than $1\times10^{-17}$ and fractional linear drift below $1\times10^{-19}$/s.
\\
\\
\end{abstract}
\pagestyle{empty}

\maketitle
\thispagestyle{empty}
Ultrastable lasers are at the core of the world's best precision measurements, including optical atomic clocks \cite{Ludlow2014, Nicholson2015}, tests of relativity \cite{Hills1990,Wiens2016}, and gravitational wave detectors \cite{Abbott2009}. 
Improved optical coherence will open the door for more precise optical clocks~\cite{Nicholson_2012,jiang_2011}.
These lasers will further studies in fundamental physics in several aspects, including the search for dark matter~\cite{stadniksisr}, atom-based gravitational wave detectors~\cite{Kolkowitz_GW} and many-body physics~\cite{martinscience}.
Furthermore, optical clocks will play a defining role in the next generation of optical timescales~\cite{riehle_si_second, ido_timescale, Riehle2018}.
All of these applications greatly benefit from improved short, mid, and long-term laser frequency stability. 

In this paper, we present critical advancements in the development of a cryogenic ultrastable optical cavity. Performance of ultrastable optical cavities is typically evaluated with respect to the fundamental thermal noise floor. 
A silicon cavity cooled to a temperature of 124~K, which corresponds to the first zero-crossing point for the silicon thermal expansion coefficient, has demonstrated a thermal noise-limited frequency stability~\cite{Matei2017}. 
Using a closed-cycle cryocooler to reach 4 K where the silicon thermal expansion asymptotically approaches zero, a cryogenic 6-cm long silicon cavity has already demonstrated fractional frequency instability of $1\times~10^{-16}$~\cite{Zhang2017}. 
With improved thermal and vibration isolation, optical feedback management, and cavity locking, we have improved the performance of this system at all averaging times. 

Short-term noise is optimized by reducing the impact of vibrations and other technical noise sources, unveiling the thermal noise floor for the first time for a 4~K optical cavity. 
Through a frequency comparison with a reference laser (named Si3)~\cite{Matei2017}, we demonstrate instability at the thermal noise floor of $6.5\times 10^{-17}$ for averaging times of 0.8 to 80 seconds. 
Furthermore, we make the discovery that the frequency drift depends linearly on the incident power. 
The drift decreases as the circulating optical power is reduced, extrapolating to a zero drift when the incident power is zero. 
The lowest drift is attained at a transmitted power of 40 nW, giving a fractional frequency drift of $-3\times~10^{-19}$/s. 
This constitutes the first demonstration of thermal noise limited performance at 4~K and the discovery of a power-dependent drift of a cryogenic optical cavity. 

\begin{figure*}
\centering
\includegraphics[width=13cm]{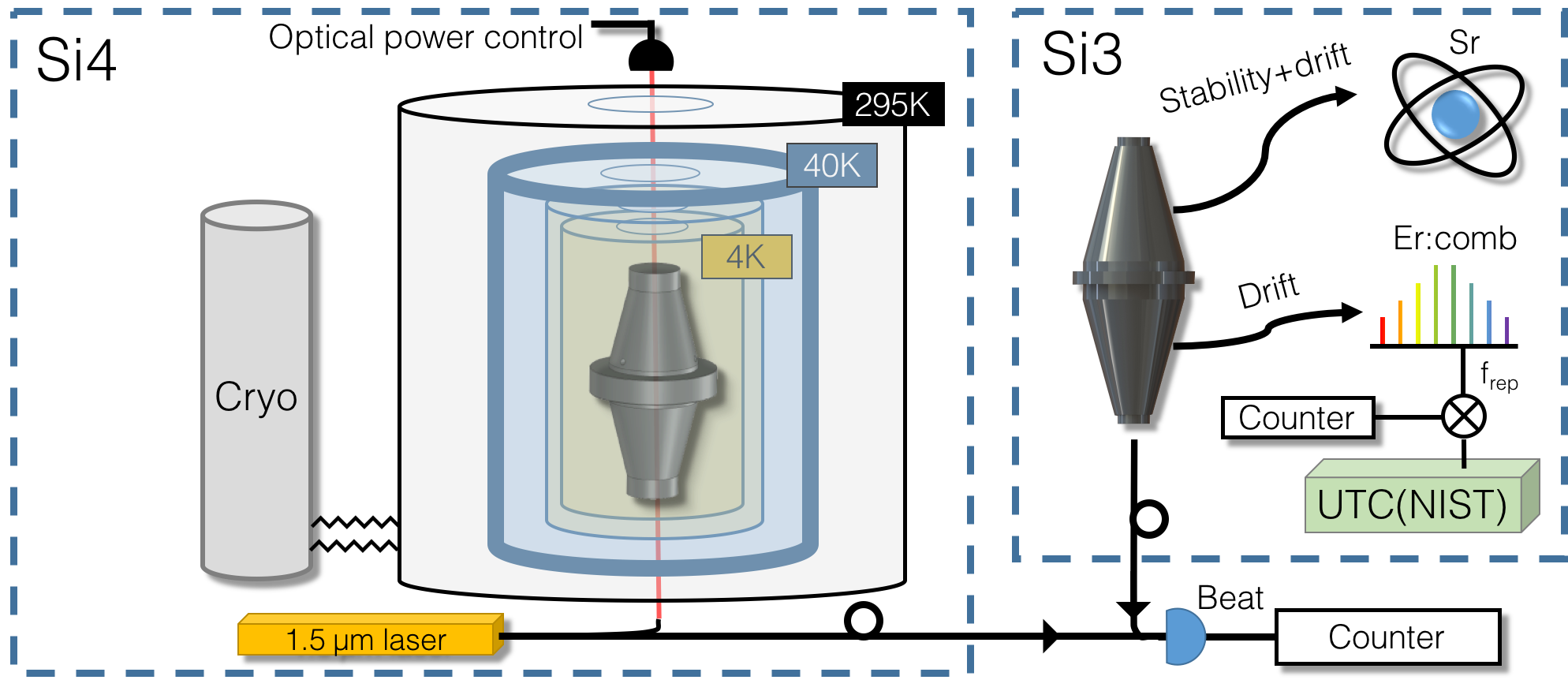}
\caption{Schematic of the optical cavity setup and measurement system. The 1.5 $\mu$m laser is stabilized to the 4 K cavity using Pound-Drever-Hall (PDH) locking. The cryostat is connected to the main chamber (black) and the cryogenic shields through a flexible vacuum bellows. The blue shield is stabilized near 40~K and encloses two inner shields near 4~K (orange). The optical power is controlled in transmission. Si3 is shown on the right, which consists of a 1.5 $\mu$m laser locked to a 124~K silicon cavity. The Si3 laser stability and drift are measured by a strontium optical lattice clock. The drift of Si3 is confirmed by comparing the repetition rate of an Er:fiber frequency comb locked to Si3 versus a hydrogen maser which is directly calibrated by the UTC(NIST) timescale.}
\label{fig:cartoon}
\end{figure*}

The schematic of the 4~K silicon cavity system (Si4) is shown in Fig.~\ref{fig:cartoon}.
The 6 cm long cavity is enclosed in a three-stage cryogenic thermal damping system formed by an outer radiation shield~(thick blue cylinder in Fig.~\ref{fig:cartoon}a), and two inner shields near 4~K~(thin orange)~\cite{Zhang2017}.
Room temperature coupling to the cavity is especially important due to the $T^4$ scaling of the radiative power.
To address this, we optimized the design of the outermost cryogenic shield.
We changed the material from aluminum to copper and added active temperature stabilization.
Copper has a thermal conductivity 100 times larger than that of aluminum at 40~K, leading to a more homogeneous and lower temperature thermal shield.
These improvements lead to a reduced coupling of room temperature variations to the cavity frequency from 200 Hz/K to 4 Hz/K.
In order to support mHz level instability, we now require only mK level control of the room temperature enclosure.

Vibrations are the primary source of short-term instability for our system. 
We minimize vibrations coming from the cryocooler by carefully designing the mechanical layout of the system~\cite{Zhang2017}. 
Due to the anisotropic nature of the silicon crystal~\cite{mateipotsdam}, we were able to reduce the vertical vibration sensitivity to $(5\pm2)\times 10^{-12}$/g at a driving frequency of 9.5 Hz.  
The horizontal vibration sensitivity in each direction was measured to be $(2\pm1)\times~10^{-10}$/g.
A reduction of vibrations at the cavity was obtained by fine tuning the relative position of the vacuum chamber and the cryostat. 
The combined improvements in sensitivity and noise provide a tenfold reduction in the frequency noise power spectral density (PSD) for Fourier frequencies of 10-50 Hz compared to previous work~\cite{Zhang2017}.

Two other critical noise sources are residual amplitude modulation (RAM) and intensity fluctuations.
We control RAM to the ppm level by employing active RAM cancellation~\cite{Zhang2014}.
Intensity fluctuations in transmission couple to the fractional frequency of the cavity with a sensitivity of $1\times 10^{-12}$/$\mu$W near DC.
To eliminate this noise source, we stabilize fluctuations in the transmitted optical power to the picowatt level by using a photodetector and feeding back to an acousto-optic modulator before the cavity. 
This both ensures that the intensity-induced frequency fluctuations are below the thermal noise floor and provides a way to change the power for investigating the cavity frequency drift.


To determine the instability of the Si4 cavity, we measure a beat between Si4 and Si3 (see Fig.~\ref{fig:cartoon}). Si3 consists of a 1.5~$\mu$m laser stabilized to a silicon cavity which operates at 124 K with a thermal noise floor of $4\times 10^{-17}$~\cite{mateipotsdam,Matei2017}. 
The short-term instability (averaging times of 0.1 to 10 s) of Si3 is determined by a three-cornered comparison with Si4 and a ULE clock laser at 698~nm.
The long-term instability (>10~s) is directly measured by a strontium optical lattice clock. These measurements show that Si3 is at its thermal noise floor for averaging times from 0.1 to 1000 s~\cite{oelker}.

\begin{figure}
\centering
\includegraphics[width=6.8cm]{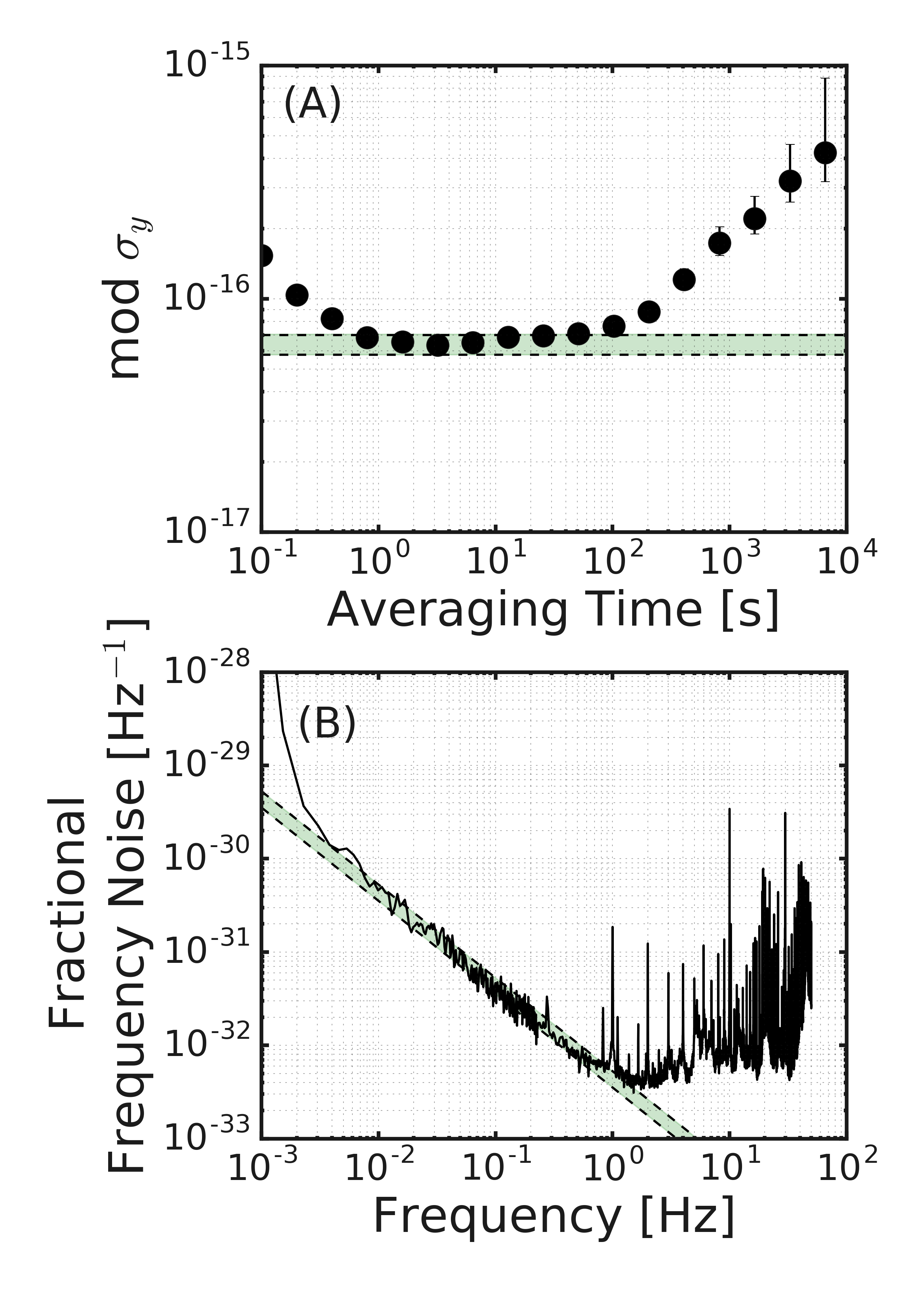}
\caption{(A) Modified Allan deviation for the Si4 cavity.
(B) Fractional frequency noise power spectral density ($S_y$) of the Si4 cavity. In both panels, the shaded green is the predicted thermal noise floor.}
\label{fig:xcorr}
\end{figure}

The modified Allan deviation of this beat, after subtracting the reference laser instability of $4\times~10^{-17}$ in quadrature, is displayed in Fig.~\ref{fig:xcorr}(A). The modified Allan deviation is calculated from a 24,000 second long measurement record made with a dead-time free lambda-type counter.
We compute the instability after removing the linear drift of the beat with a magnitude of $\sim 3\times 10^{-18}$/s.
The Si4 instability reaches $6.5 \times10^{-17}$ for averaging times of $0.8< \tau<80$ s, which is consistent with the predicted thermal noise floor~(green shaded region). 
The uncertainty in the thermal noise floor arises from the spread in the loss angle at 4 K~\cite{yamamoto_coating,hirose_coating}. 

The corresponding frequency noise PSD for Si4 is shown in Fig.~\ref{fig:xcorr}(B). 
The PSD is calculated from the time-series of the beat obtained by the frequency counter.
The thermal noise floor of the Si3 cavity~($S_y = 1.7\times10^{-33}/f$) is subtracted from the beat PSD~\cite{Matei2017}. 
The Si4 laser is limited by the thermal noise floor for Fourier frequencies over nearly three decades, from 5 mHz to 2 Hz.
We fit the measured PSD to a function $S_y = a f^{-1}$ and obtain the fit parameter $a=4.12(5)~\times~10^{-33}$. 
Using the full expression given in~\cite{Cole2013}, the Young's modulus and Poisson's ratio from~\cite{coatingyoungs}, we extract a loss angle for the SiO$_2$/Ta$_2$O$_5$ mirror coatings to be $\phi = 5.1(5) \times 10^{-4}$. 
The thin noise spikes at 1 Hz and higher harmonics come from the cryocooler vibrations. The laser deviates from thermal noise at Fourier frequencies below 0.5 mHz, potentially due to etalons or temperature fluctuations.

\begin{figure}[h]
\centering
\includegraphics[width=6.2cm]{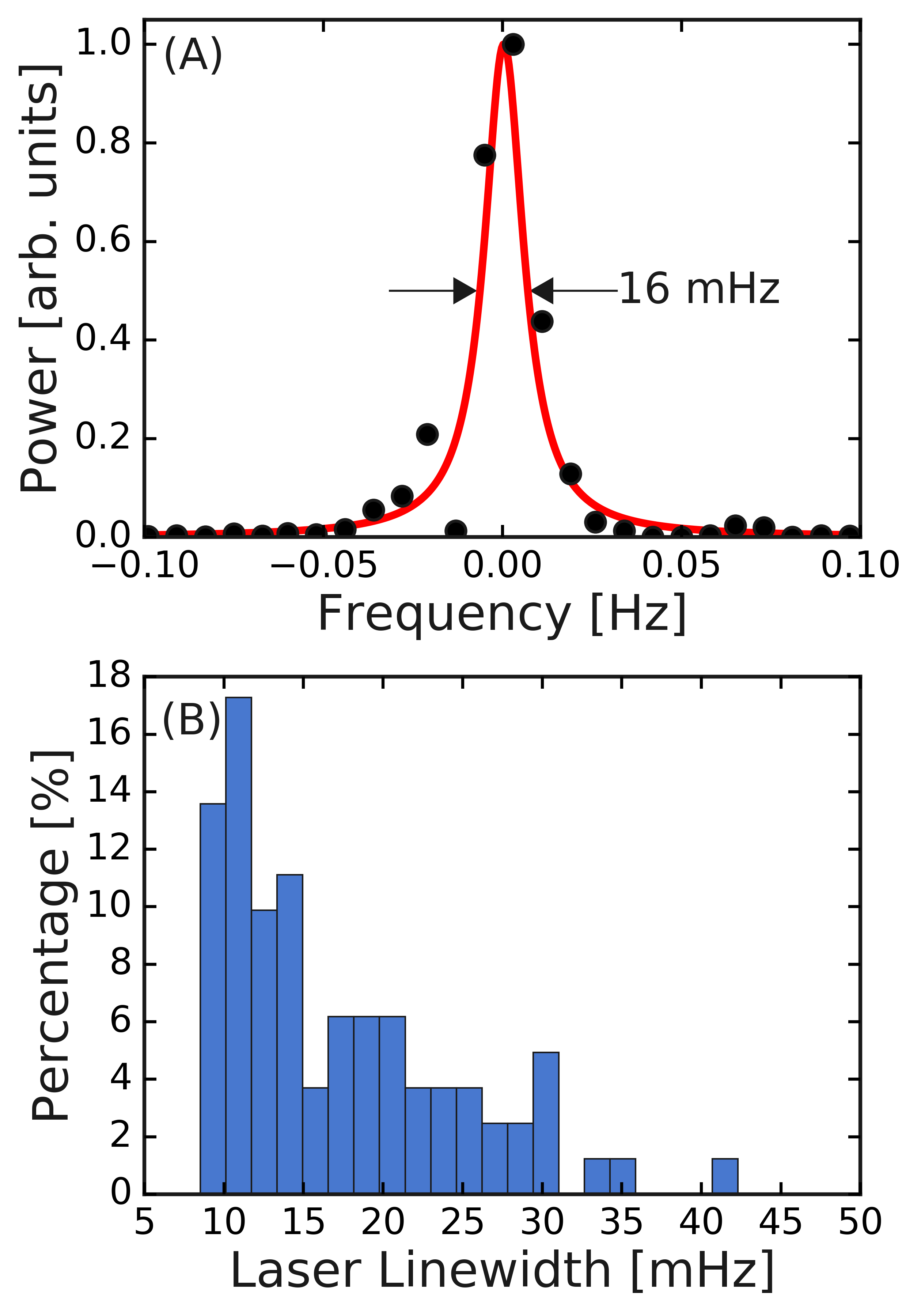}
\caption{(A) FFT of the beat measured at 1542~nm (black circles), fit to a Lorentzian lineshape (red line). (B) Histogram of the measured Si4 linewidths for 100 measurements. 
}
\label{fig:linewidth}
\end{figure}


We experimentally determine the laser linewidth from a Fast Fourier Transform (FFT) of the Si3-Si4 beat. 
The beat is mixed down to 10 Hz and digitized with a analog-to-digital converter.
One example of such a measurement is shown in Fig.~\ref{fig:linewidth}(A). 
We use a measurement time of 128 seconds and employ a Hanning window, corresponding to a Fourier limit of 10.9~mHz.
The expected linewidth for $1/f$ frequency noise is given by a statistical distribution~\cite{Matei2017}.
The distribution is multiplied by the ratio $\sigma_{Si4}/(\sqrt{\sigma_{Si3}^2 + \sigma_{Si4}^2}) = 0.85$ in order to estimate the relative contribution of the Si4 cavity to the beat linewidth. 
Here, $\sigma_{Si4(Si3)}$ refers to the thermal noise floor of Si4(Si3).
We repeat this measurement 100 times and plot the histogram of results in Fig.~\ref{fig:linewidth}(B). 
The median laser linewidth for the distribution in Fig.~\ref{fig:linewidth}(B) is 16 mHz, which represents the lowest observed to date for an optical cavity placed inside a closed-cycle cryocooler.

We measure the drift of the Si4 system by counting the beat Si3-Si4 as shown in Fig.~\ref{fig:cartoon}.
This requires careful calibration of the drift of the Si3 system.
The Si3 laser is used as the clock laser for a strontium optical lattice clock, giving a direct measurement of the drift~\cite{oelker}.
As an independent check, the drift of the Si3 system is continuously monitored against a hydrogen maser from NIST via an optical frequency comb.
This maser is then calibrated against UTC(NIST) as depicted in Fig.~\ref{fig:cartoon}.
The long-term linear frequency drift of Si3 is $-3\times~10^{-19}$/s with 2.8~$\mu$W of transmitted power.
The measured linear drift of Si3 is removed from the Si3-Si4 beat, thus giving the drift of Si4.

\begin{figure*}
\centering
\includegraphics[width=14cm]{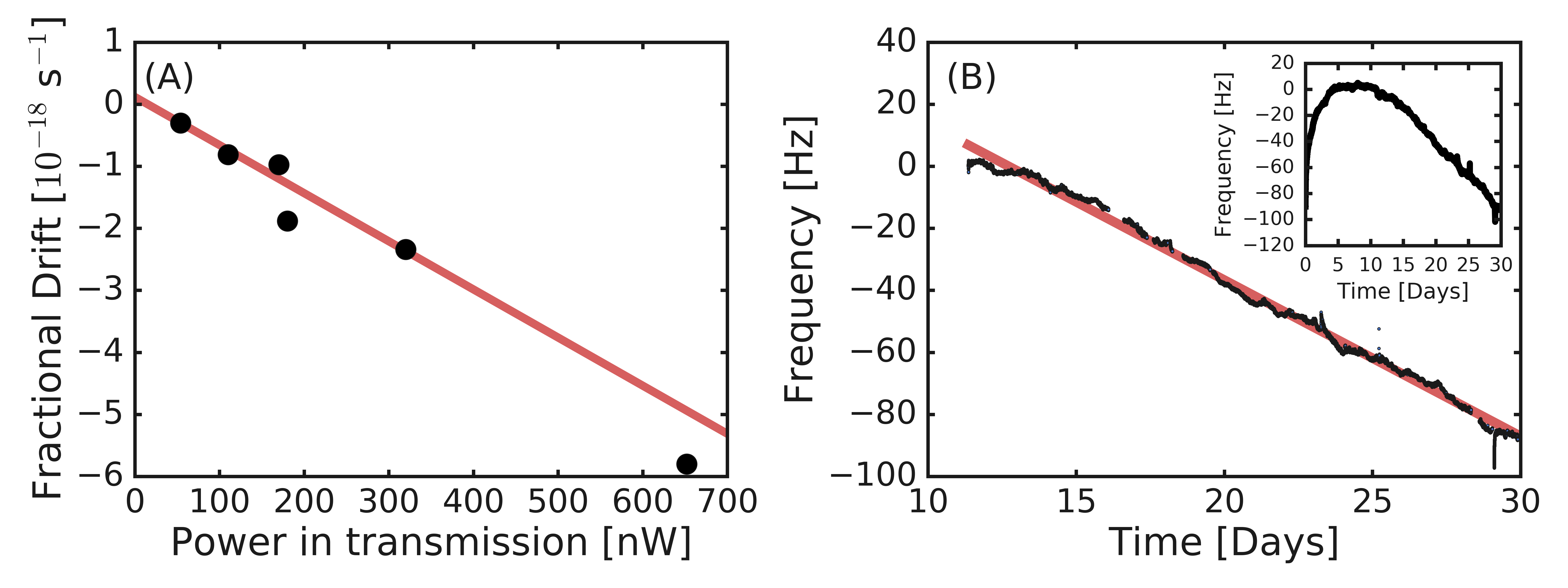}
\caption{(A) Fractional frequency drift of Si4 as a function of optical power in transmission. The red line is a linear fit to the data, shown to guide the eye. (B) Optical frequency of Si4 with 40~nW in transmission, corresponding to the lowest optical power measured. The red line is a linear fit to the data, corresponding to a drift rate of $-3\times 10^{-19}$/s. The inset shows the complete frequency record, where $t=0$ corresponds to the time when the laser is locked to the cavity.}
\label{fig:drift}
\end{figure*}

The linear frequency drift of the Si4 cavity is dependent on the transmitted optical power as shown in Fig.~\ref{fig:drift}(A).
We vary the incident power and stabilize the cavity transmission at various levels as shown in Fig.~\ref{fig:cartoon}.
With a cavity finesse of $F = 500,000$ and a transmission coefficient of $T=2$~ppm, a transmitted power of 40~nW corresponds to a circulating optical power of 2~mW.
Each time the optical power in the cavity is changed, a frequency transient is observed with a characteristic time constant between 1 and 2 days.
In order to extract the linear frequency drift, we typically wait several time constants for the transient to decay away.
When the drift is low (at lower optical power), we wait even longer in order to avoid the contribution from the transient.
For example, as shown in the inset of Fig.~\ref{fig:drift}(B), we wait 5 time constants before fitting a linear drift.
We achieve high performance at low optical power by employing resonant photodetectors for both the PDH and the RAM detection, providing a shot-noise limited signal-to-noise ratio at 68 nW. 

The linear power dependence of the drift is striking evidence for a new mechanism of length drift of an optical cavity at low temperatures. 
The sign of the frequency drift is always negative, meaning the cavity is getting longer over time. 
The slope of the power dependence is roughly $-7\times~10^{-21}$/s/nW. 
One potential explanation is thermal-induced mechanical creep of the mirror coating, where the mismatch in the coefficient of thermal expansion for the substrate and the coating gives a temperature-dependent creep.
To reduce the impact of optical power on the long-term drift, Wiens et al. minimized the irradiation of their mirrors by periodically scanning the laser across the cavity resonance to measure the cavity frequency~\cite{Wiens2016}.
We present the first rigorous characterization of a power-dependent frequency drift in an optical cavity.

The lowest operating power we have achieved is 40~nW in transmission, giving a fractional frequency drift of $-3\times 10^{-19}$/s.
This frequency drift is comparable to the previous state-of-the-art obtained from a 124 K silicon cavity~\cite{Hagemann_OL}.
However, the implication of the current finding is tantalizing in that as we continue to reduce the incident power, we can access an extremely low value of cavity drift, making it possible that such a cavity alone could be useful as a potential time scale.  
At this low power, the fractional noise of the laser is higher than that showed in Fig.~\ref{fig:xcorr} by about a factor of two. The extra noise is due to the photodetector, and will be mitigated with an improved design.

The advances presented here point to a clear direction for ultrastable lasers.
To operate with a minimal frequency drift, optical cavities at low temperature will have to operate at very low optical power.
Reduction of the thermal noise will be possible by replacing the conventional SiO$_2$/Ta$_2$O$_5$ mirrors with crystalline mirrors~\cite{Cole2016,Cole2013}.
Such crystalline mirrors have been shown to exhibit a factor of 10 lower loss angle at room temperature~\cite{Cole2013}. 
Increasing the cavity length further will also reduce the fractional frequency noise.
We can now foresee a strong possibility of achieving an ultrastable cavity with fractional instability $< 1 \times 10^{-17}$ using a continuously-running closed-cycle cryocooler at 4 K.\\
\\
\textbf{Acknowledgments} The authors thank T. Brown and T. Asnicar for technical assistance.
We acknowledge technical contributions from T. Bothwell, D. Kedar, C. Kennedy, C. Sanner and L. Sonderhouse.
We thank J. Sherman for the UTC(NIST) timescale data.
\\

\noindent\textbf{Funding} This work is supported by NIST, DARPA, JILA
Physics Frontier Center (NSF PHY-1734006), Centre for Quantum Engineering and SpaceTime Research (QUEST), and Physikalisch-Technische Bundesanstalt. 
T. L., D. G. M. and U. S. acknowledge support  from the Quantum sensors (Q-SENSE) project, supported by the European Commission’s H2020 MSCA RISE under Grant Agreement Number 69115.
E. O. is supported by the National Research
Council postdoctoral fellowship.

\bibliography{si4bib_new}




\end{document}